\documentclass[letterpaper,prc,preprint]{revtex4}
\usepackage[dvips]{graphicx}
\usepackage{amssymb, latexsym, amsmath, amsfonts, fullpage}
\usepackage{hyperref}
\usepackage{tabularx}
\usepackage{url}

\begin{document}

\title{Symmetry energy effects on location of the inner edge of neutron star crusts}

\author{Ch C Moustakidis$^{1,2}$}
\email{moustaki@auth.gr}
\address{$^{1}$Department of Theoretical Physics, Aristotle University of
Thessaloniki,  54124 Thessaloniki, Greece \\
$^{2}$Theoretical Astrophysics, University of Tuebingen IAAT,
Auf der Morgenstelle 10, Tuebingen 72076 Germany}

\begin{abstract}
The symmetry energy effects on the location of the inner edge of
neutron star crusts are studied. Three phenomenological models are
employed in order to check the accuracy of the well known
parabolic approximation of the equation of state for asymmetric
nuclear matter in the determination of the transition density
$n_t$ and transition pressure $P_t$. The results corroborate the
statement that the error due to the assumption that  a priori the
equation of state is parabolic may introduce a large error in the
determination of related properties of a neutron star as the
crustal fraction of the moment of inertia and the critical
frequency of rotating neutron stars.

\vspace{0.3cm}

PACS number(s): 21.65.cd, 21.65.Ef, 21.65.Mn, 26.60.Kp,
26.60.-c, 26.60.Gj. \\

Keywards: Equation of state; nuclear symmetry energy; neutron star
crust.
\end{abstract}

\maketitle
\newpage

\section{Introduction}
Neutron stars (NS) are extraordinary astronomical laboratories for
the physics of dense neutron-rich nuclear matter
\cite{Shapiro-83,Haensel-07}. The main parts of a NS are the crust
and the core. The latter, divided into the outer core and the
inner one, has a radius of approximately 10 km and contains most
of the star's mass while the crust, with a thickness of about $1$
km and containing only a few percent of the total mass, can also
be divided into an outer and an inner part. A very important
ingredient in the study of the structure and various properties of
neutron stars is the equation of state (EOS) of neutron-rich
nuclear matter~\cite{Lattimer-07}.

One of the most important predictions of a given EOS is the
location of the inner edge of a neutron star crust. The inner
crust comprises the outer region from the density at which
neutrons drip-out of nuclei, to the inner edge separating the
solid crust from the homogeneous liquid core. At the inner edge,
in fact, a phase transition occurs from the high-density
homogeneous matter to the inhomogeneous one at lower densities.
The transition density takes its critical value $n_c$ when the
uniform neutron-proton-electron matter (npe) becomes unstable with
respect to the separation into two coexisting phases (one
corresponding to nuclei, the other one  to a nucleonic sea)
\cite{Lattimer-07}.

While the density at which neutrons drip-out of nuclei is rather
well determined, the transition density $n_t$ at the inner edge is
much less certain due to our insufficient knowledge of the EOS of
neutron-rich nuclear matter. The value of $n_t$ determines the
structure of the inner part of the crust. If sufficiently high, it
is possible for non-spherical phases, with rod- or plate-like
nuclei, to occur before the nuclei dissolve. If $n_t$ is
relatively low, then the matter undergoes a direct transition from
spherical nuclei to uniform nucleonic fluid. The extent to which
non-spherical phases occur will have important consequences for
other properties  determined by the solid
crust~\cite{Pethick-95b}.

In general, the determination of the transition density $n_t$
itself is a very complicated problem because the inner crust may
have a very complicated structure. A well established approach is
to find the density at which the uniform liquid first becomes
unstable against small-amplitude density fluctuations, indicating
the formation of nuclear clusters. This approach includes the
dynamical method
\cite{Pethick-95b,Baym-71a,Baym-71b,Pethick-95a,Douchin-00,Oyamatsu-07,Ducoin-07,Xu-09-1,Xu-09-2},
the thermodynamical one
\cite{Lattimer-07,Kubis-07,Worley-08,Kubis-04}, and the random
phase approximation (RPA)~\cite{Horowitz-01,Carriere-03}.

Theoretical studies have shown that the core-crust transition
density and pressure are very sensitive to the density dependence
of the nuclear matter symmetry
energy~\cite{Lattimer-07,Xu-09-1,Xu-09-2,Kubis-07,Worley-08,Kubis-04,Horowitz-01,
Carriere-03,Ducoin-011,Ducoin-010,Moustakidis-010,Xu-010,Cai-012}.
At present, the symmetry energy is well constrained experimentally
up to the value of the nuclear saturation density $n_0$ but still
remains almost unknown in the density regime appropriate for the
interior of neutron stars ($n\gg n_0$)
~\cite{LCK08,Tsa04,Tsang-09}. In spite of the experimental
uncertainty  of the symmetry energy, there are many theoretical
considerations of the symmetry energy categorized mainly in
phenomenological and effective field theoretical approaches. The
aim  of this work is: a) to apply a momentum dependent interaction
model (MDI) as well as  two additional non relativistic models
(Thomas-Fermi and Skyrme type)  to determine the transition
density and transition pressure  corresponding to the edge of a
neutron star crust and b) to check the accuracy of the parabolic
approximation  widely used in the literature and applied  to
neutron star research.

The article is organized as follows. In Sec.~2 we review the
Taylor expansion of the energy while in Sec.~3 we present the
thermodynamical method for the determination of the transition
density and pressure of the inner edge of a neutron star crust.
The nuclear models  employed in the the present work  are
presented in Sec.~4. The results  are presented and discussed in
Sec.~5. Finally Sec.~6 summarizes the present study.

\section{Taylor expansion of the energy}

The energy per particle $E(n,I)$ in cold asymmetric nuclear matter
can be expanded around $I=0$ as follows
\begin{equation}
E(n,I)=E(n,I=0)+E_{sym,2}(n)I^2+E_{sym,4}(n)I^4+\dots
+E_{sym,2k}(n)I^{2k}+\dots, \label{Expans-1}
\end{equation}
where the total baryon density $n=n_p+n_n$, $n_p$ ($n_n$) is the
proton (neutron) density,  $I=(n_n-n_p)/n$ is the asymmetry
parameter and  $E(n,I=0)$ is the energy per baryon of the
symmetric nuclear matter, while  the coefficients of the expansion
are
\begin{eqnarray}
&&E_{sym,2}(n)=\left. \frac{1}{2!}\frac{\partial^2E(n,I)}{\partial
I^2} \right|_{I=0}, \quad E_{sym,4}(n)=\left.
\frac{1}{4!}\frac{\partial^4E(n,I)}{\partial I^4}\right|_{I=0},
\nonumber \\
&& E_{sym,2k}(n)=\left.
\frac{1}{(2k)!}\frac{\partial^{2k}E(n,I)}{\partial
I^{2k}}\right|_{I=0}. \label{Expan-2}
\end{eqnarray}
In (\ref{Expans-1}), only even powers of $I$ appear due to the
fact that the strong interaction must be symmetric under exchange
of  neutrons with protons i.e.  the contribution to the energy
must be independent of the sign of the difference $n_n-n_p$.
The second order approximation of the expansion (\ref{Expans-1})
is written as
\begin{equation}
E(n,I)\simeq E(n,I=0)+E_{sym,2}(n)I^2. \label{Expans-3}
\end{equation}
We expect, from a mathematical point of view, that the above
expansion is accurate at least close to $I=0$ (the case of
symmetric nuclear matter). However for the majority of the energy
functionals, the above approximation works well   for higher
values of the asymmetry parameter $I $  and even  close to the
value $I=1$ corresponding to the pure neutron matter. Thus, in
cases for which the expansion, in a good approximation, is
independent of the asymmetry parameter $I$, the symmetry energy
can be defined as
\begin{equation}
E_{sym}(n)=E(n,I=1)-E(n,I=0), \label{esym-def}
\end{equation}
and the energy per baryon is written
\begin{equation}
E(n,I)=E(n,I=0)+\underbrace{\left(E(n,I=1)-E(n,I=0)\right)}_{E_{sym}(n)}I^2.
\label{Par-1}
\end{equation}
The knowledge of the equation of state of neutron rich matter is
fundamental in astrophysical applications. For example it is the
basic ingredient for the study of $\beta$-stable matter
characteristic for the interior of neutron stars. Actually, for
the most of the equations of state (mainly those originating from
microscopic calculations or those coming from relativistic mean
field theories)  only the energy of symmetric nuclear matter and
pure neutron matter are determined and the definition of the
symmetry energy from equation (\ref{esym-def}) is almost
unavoidable. The question  naturally arising is the magnitude of
the width of the error  introduced by assuming a priori that the
equation of state is parabolic according to relations
(\ref{Expans-1}) and (\ref{Par-1}).

In the present work we define in addition the {\it second order
expansion} of the form
\begin{equation}
E(n,I)\simeq E(n,I=0)+E_{sym,2}(n)I^2=E(n,I=0)+\left.
\frac{1}{2!}\frac{\partial^2E(n,I)}{\partial I^2}
\right|_{I=0}I^2, \label{Second-order-1}
\end{equation}
which is similar to expansion (\ref{Par-1})   replacing the
quantity $E_{sym}(n)=E(n,I=1)-E(n,I=0)$ by $\displaystyle
E_{sym,2}(n)=\left. \frac{1}{2!}\frac{\partial^2E(n,I)}{\partial
I^2} \right|_{I=0}$.

Before trying to check the accuracy of the approximation
(\ref{Par-1}) it is worth to compare the density dependence of the
above two definitions of the symmetry energy i.e.  $E_{sym,2}(n)$
and $E_{sym}(n)$ given by  (\ref{Expan-2}) and (\ref{esym-def})
respectively.

\section{The thermodynamical method}
The core-crust interface corresponds to the phase transition
between nuclei and uniform nuclear matter. The uniform matter is
nearly pure neutron matter, with a proton fraction of just a few
percent determined by the condition of beta equilibrium. Weak
interactions conserve both baryon number and charge
\cite{Lattimer-07}, and from the first law of thermodynamics, at
temperature $T=0$ we have
\begin{equation}
{\rm d}u=-P{\rm d}v-\hat{\mu}{\rm d}q, \label{u-1}
\end{equation}
where $u$ is the internal energy per baryon, $P$ is the total
pressure,  $v$ is the volume per baryon ( $v=1/n$ where $n$ is the
baryon density) and $q$ is the charge fraction ($q=x-Y_e$ where
$x$ and $Y_e$ are the proton and electron  fractions in baryonic
matter respectively). In $\beta$-equilibrium the chemical
potential $\hat{\mu}$ is given by $\hat{\mu}=\mu_n-\mu_p=\mu_e$
where $\mu_p$, $\mu_n$ and $\mu_e$  are the chemical potentials of
the protons, neutrons and electrons respectively.  The stability
of the uniform phase requires that $u(v,q)$ is a convex
function~\cite{Callen-85}. This condition leads to the following
two constraints for the pressure and the chemical potential
\begin{equation}
-\left(\frac{\partial P}{\partial v}\right)_q-\left(\frac{\partial
P}{\partial q}\right)_v \left(\frac{\partial q}{\partial
v}\right)_{\hat{\mu}}>0, \label{cond-1}
\end{equation}
\begin{equation}
-\left(\frac{\partial \hat{\mu}}{\partial q}\right)_v>0.
\label{cond-2}
\end{equation}
It is assumed that the total internal energy per baryon $u(v,q)$
can be decomposed into baryon ($E_N$) and electron ($E_e$)
contributions
\begin{equation}
u(v,q)=E_N(v,q)+E_e(v,q). \label{u-2}
\end{equation}
The related theory has been extensively presented in our recent
publication~\cite{Moustakidis-010}. We consider the condition of
charge neutrality $q=0$ which requires that $x=Y_e$. This is the
case we will consider also  in the present study. Hence, according
to Ref.~\cite{Moustakidis-010} the constraints (\ref{cond-1}) and
(\ref{cond-2}), in the case of the full EOS (FEOS) and the
parabolic approximation (PA), are written
\begin{equation}
C_{{\rm FEOS}}(n)=2n\frac{\partial E(n,x)}{\partial
n}+n^2\frac{\partial ^2E(n,x)}{\partial
n^2}-\left(\frac{\partial^2E(n,x)}{\partial n
\partial x}n \right)^2\left(\frac{\partial^2E(n,x)}{\partial x^2}
\right)^{-1} > 0,  \label{cont-1}
\end{equation}
\begin{eqnarray}
C_{{\rm PA}}(n)&=&n^2\frac{{\rm d}^2E(n,x=0.5)}{{\rm
d}n^2}+2n\frac{{\rm d}E(n,x=0.5)}{dn}+(1-2x)^2 \label{K-I}\\
&\times& \left[ n^2 \frac{{\rm d}^2 E_{sym}(n)}{{\rm
d}n^2}+2n\frac{{\rm d}E_{sym}(n)}{{\rm
d}n}-2\frac{1}{E_{sym}(n)}\left(n \frac{{\rm d}E_{sym}(n)}{{\rm
d}n} \right)^2 \right]>0. \nonumber
\end{eqnarray}
For a given equation of state, the quantity $C_{{\rm FEOS}}(n)$
(or $C_{{\rm PA}}(n)$) is plotted as a function of the baryonic
density $n$ and the equation $C_{{\rm FEOS}}(n)=0$ (or $C_{{\rm
PA}}(n)=0$) defines the transition density $n_t$. However, what
remains is the determination of the proton fraction $x$ (as a
function of the baryon density $n$) in $\beta$-stable matter. In
this case we have the processes
\begin{equation}
n \longrightarrow p+e^{-}+\bar{\nu}_e \qquad \qquad p +e^{-}
\longrightarrow n+ \nu_e
\end{equation}
which take place simultaneously. We assume that neutrinos
generated in these reactions have left the system. This implies
that
\begin{equation}
\hat{\mu}=\mu_n-\mu_p=\mu_e ,\label{chem-1}
\end{equation}
Given the total energy density of the baryons $\epsilon_N \equiv
\epsilon(n_n,n_p)$, the neutron and proton chemical potentials can
be defined as (see also Ref.~\cite{Prakash-94})
\begin{equation}
\mu_n=\left(\frac{\partial \epsilon_N}{\partial n_n}\right)_{n_p},
\qquad \qquad \mu_p=\left(\frac{\partial \epsilon_N }{\partial
n_p}\right)_{n_n} . \label{chem-2}
\end{equation}
It is easy to show that after some algebra we get
\begin{equation}
\hat{\mu}=\mu_n-\mu_p=-\left(\frac{\partial (\epsilon_N
/n)}{\partial x}\right)_n=\left( -\frac{\partial E_N}{\partial
x}\right)_n . \label{chem-3}
\end{equation}
The charge condition implies that $n_e=n_p=nx$ or
$k_{F_e}=k_{F_p}$ (where $k_F$ are the fermi momenta). In
addition, the chemical potential of the electron is given by the
relation (relativistic electrons)
\begin{equation}\mu_e=\sqrt{k_{F_e}^2c^2+m_e^2c^4}\simeq k_{F_e}
c=\hbar c(3 \pi^2 n x)^{1/3}. \label{chem-ele-1}
\end{equation}
Finally, from equations (\ref{chem-3}) and (\ref{chem-ele-1}) one
has
\begin{equation}
\left( \frac{\partial E_N}{\partial x}\right)_n=-\hbar c(3 \pi^2 n
x)^{1/3}. \label{x-frac}
\end{equation}
Equation (\ref{x-frac}) is the most general relation that
determines the proton fraction of $\beta$-stable matter. In the
case of the PA the above equation, with the help of
Eq.~(\ref{Par-1}) is written as
\begin{equation}
4(1-2x)E_{sym}(n)=\hbar c(3 \pi^2 n_e)^{1/3}=\hbar c(3 \pi^2 n
x)^{1/3}.  \label{b-equil-2}
\end{equation}
The pressure $P_t$ at the inner edge is an important quantity
directly related to the crustal fraction of the moment of inertia,
which can be measured indirectly from observations of pulsars
glitches~\cite{Lattimer-07}. The total pressure is decomposed also
into baryon and lepton contributions
\begin{equation}
P(n,x)=P_N(n,x)+P_e(n,x), \label{P-all-1}
\end{equation}
where
\begin{equation}
P_N(n,x)=n^2\frac{\partial E_N}{\partial  n}. \label{Pb-1}
\end{equation}
The electrons are considered as a non-interacting Fermi gas. Their
contribution to the total pressure reads
\begin{equation}
P_e(n,x)=\frac{1}{12\pi^2}\frac{\mu_e^4}{(\hbar c)^3}=\frac{\hbar
c}{12 \pi^2}\left(3\pi^2 xn\right)^{4/3}. \label{Pe-2}
\end{equation}
The transition pressure, in the case of the FEOS, is given now by
the equation
\begin{equation}
P_t^{FEOS}(n_t,x_t)=n_t^2\left.\frac{\partial E_N}{\partial
n}\right|_{n=n_t}+\frac{\hbar c}{12 \pi^2}\left(3\pi^2
x_tn_t\right)^{4/3}, \label{Pr-tra}
\end{equation}
where $x_t$ is the proton fraction related to the transition
density. In the case of the PA  $P_t$ is given by the relation
\begin{eqnarray}
P_t^{PA}(n_t,x_t)&=&n_t^2\left(\left. \frac{{\rm d}
E(n,x=0.5)}{{\rm d} n}\right|_{n=n_t}+\left.\frac{{\rm d}
E_{sym}(n)}{{\rm d} n}\right|_{n=n_t}(1-2x_t)^2 \right)\nonumber
\\
&+&\frac{\hbar c}{12 \pi^2}\left(3\pi^2 x_tn_t\right)^{4/3}.
\label{Pr-tra-para}
\end{eqnarray}

\subsection{Application I: crustal fraction of the moment of
inertia}
The crustal fraction of the moment of inertia $\Delta I/I$ can be
expressed as a function of $M$ (star's total mass)  and $R$
(star's radius) with the only dependence on the equation of state
arising from the values of $P_t$ and $n_t$. Actually, the major
dependence is on the value of $P_t$, since $n_t$ enters only as a
correction according to the following approximate
formula~\cite{Link-99}
\begin{equation}
\frac{\Delta I}{I}\simeq \frac{28\pi P_t
R^3}{3Mc^2}\frac{(1-1.67\beta-0.6\beta^2)}{\beta}\left(1+\frac{2P_t}{n_tmc^2}\frac{(1+7\beta)(1-2\beta)}{\beta^2}
\right)^{-1},  \label{inertia-1}
\end{equation}
where $\beta=GM/Rc^2$. The crustal fraction of the moment of
inertia is particularly interesting as it can be inferred from
observations of pulsar glitches, the occasional disruptions of the
otherwise extremely regular pulsations from magnetized, rotating
neutron stars~\cite{Xu-09-2}. Link et al.~\cite{Link-99} showed
that glitches represent a self-regulating instability for which
the star prepares over a waiting time. The angular momentum
requirements of glitches in the Vela pulsar indicate that more
than $0.014$ of the star's moment of inertia drives these events.
So, if glitches originate in the liquid of the inner crust, this
means that $\Delta I/I>0.014$

\subsection{Application II: r-mode instability of rotating neutron star}
The r-modes are oscillations of rotating stars whose restoring
force is the Coriolis force
\cite{Lidblom-2000,Andersson-1998,Friedman-98,Friedman-99,Andersson-2001,Andersson-2003,Kokkotas-99,Andersson-99,Bildsten-2000,Wen-012,Vidana-012,Alford-2012}.
The gravitational radiation-driven instability of these modes has
been proposed as an explanation for the observed relatively low
spin frequencies of young neutron stars and of accreting neutron
stars in low-mass X-ray binaries as well. This instability can
only occur when the gravitational-radiation driving time scale of
the r-mode is shorter than the time scales of the various
dissipation mechanisms that may occur in the interior of the
neutron star.

The nuclear EOS affects the time scales associated with the
r-mode, in two different ways. Firstly, EOS defines the radial
dependence of the mass density distribution $\rho(r)$, which is
the basic ingredient of the relevant integrals. Secondly, it
defines the core-crust transition density $\rho_c$ and also the
core radius $R_c$ which is the upper limit of the mentioned
integrals.

The critical angular velocity $\Omega_c$, above which the {\it
r}-mode is unstable, for $m=2$ is given  by \cite{Lidblom-2000}
\begin{equation}
\frac{\Omega_c}{\Omega_0}=\left(\frac{\tilde{\tau}_{GR}}{\tilde{\tau}_v}
\right)^{2/11}\left(\frac{10^8 \ K}{T}  \right)^{2/11}.
\label{Omega-c-1}
\end{equation}
where $\Omega_0=\sqrt{\pi G\overline{\rho}}$ and
$\overline{\rho}=3M/4\pi R^3$ is the mean density of the star and
$T$ is the temperature. $\tilde{\tau}_{GR}$ and $\tilde{\tau}_v$
are the fiducial gravitational radiation time scale and the
fiducial viscous time scale respectively given by
\begin{equation}
\tau_{GR}=\tilde{\tau}_{GR}\left(\frac{\Omega_0}{\Omega}
\right)^{2m+2},  \label{fid-t-GR}
\end{equation}
\begin{equation}
\tau_v=\tilde{\tau}_v\left(\frac{\Omega_0}{\Omega} \right)^{1/2}
\left(\frac{T}{10^8 \ K} \right), \label{fid-t-v}
\end{equation}
where \cite{Lidblom-2000}
\begin{equation}
\frac{1}{\tau_{GR}}=-\frac{32\pi G
\Omega^{2m+2}}{c^{2m+3}}\frac{(m-1)^{2m}}{[(2m+1)!!]^2}\left(\frac{m+2}{m+1}\right)^{2m+2}
\int_0^{R_c}\rho(r) r^{2m+2} dr, \label{tgr-1}
\end{equation}
and
\begin{equation}
\tau_{v}=\frac{1}{2\Omega}\frac{2^{m+3/2}(m+1)!}{m(2m+1)!! {\cal
I}_m}\sqrt{\frac{2\Omega R_c^2\rho_c}{\eta_c}}
\int_0^{R_c}\frac{\rho(r)}{\rho_c}\left(\frac{r}{R_c}\right)^{2m+2}
\frac{dr}{R_c}.
 \label{tv-1}
\end{equation}
$\Omega$ is the angular velocity of the unperturbed star,
$\rho(r)$ is the radial dependence of the mass density of the
neutron star, $R_c$, $\rho_c$ and $\eta_c$ are the radius, density
and viscosity of the fluid at the outer edge of the core. In the
present work we consider the case of $m=2$ r-mode.

\section{The models}
In the present work we employ three different phenomenological
models for  the energy per baryon  of the asymmetric nuclear
matter having the advantage of an analytical form. The MDI model
has been extensively applied for  neutron star studies, can
reproduce the results of more microscopic calculations of dense
matter at zero temperature and can be extended to finite
temperature~\cite{Prakash-97,Moustakidis-07-1,Moustakidis-07-2,Moustakidis-08,Moustakidis-09-1,Moustakidis-09-2}.
The Skyrme model using various parametrizations can also be
applied  both in nuclear matter and in finite
nuclei~\cite{Chabanat-97,Farine-97}. Finally we employ  a version
of the Thomas-Fermi model, which was introduced by Myers {\it et
al}.~\cite{Myers-98}  and has also been applied for the study of
finite nuclei and in a few cases for high density nuclear matter
applications~\cite{Strobel-97}.

\subsection{MDI model}

The model used here, which has already been presented and analyzed
in previous papers
\cite{Prakash-97,Moustakidis-07-1,Moustakidis-07-2,Moustakidis-08,Moustakidis-09-1,Moustakidis-09-2},
is designed to reproduce the results of the microscopic
calculations of both nuclear and neutron-rich matter at zero
temperature and can be extended to finite temperature
\cite{Prakash-97}. The energy density  of the asymmetric nuclear
matter (ANM), in MDI model is given by the relation
\begin{equation}
\epsilon(n_n,n_p,T=0)=\epsilon_{kin}^{n}(n_n,T=0)+\epsilon_{kin}^{p}(n_p,T=0)+
V_{int}(n_n,n_p,T=0), \label{E-D-1}
\end{equation}
where the first two terms are the kinetic energy contributions on
the total energy density, while the third one is the potential
energy contribution. The energy per baryon  at $T=0$, is given by
\begin{eqnarray}
E(n,I)&=&\frac{3}{10}E_F^0u^{2/3}\left[(1+I)^{5/3}+(1-I)^{5/3}\right]+
\frac{1}{3}A\left[\frac{3}{2}-(\frac{1}{2}+x_0)I^2\right]u
\nonumber \\ &+&
\frac{\frac{2}{3}B\left[\frac{3}{2}-(\frac{1}{2}+x_3)I^2\right]u^{\sigma}}
{1+\frac{2}{3}B'\left[\frac{3}{2}-(\frac{1}{2}+x_3)I^2\right]u^{\sigma-1}}
 \label{e-T0}\\
&+&\frac{3}{2}\sum_{i=1,2}\left[C_i+\frac{C_i-8Z_i}{5}I\right]\left(\frac{\Lambda_i}{k_F^0}\right)^3
\left(\frac{\left((1+I)u\right)^{1/3}}{\frac{\Lambda_i}{k_F^0}}-
\tan^{-1} \frac{\left((1+
I)u\right)^{1/3}}{\frac{\Lambda_i}{k_F^0}}\right)\nonumber \\
&+&
\frac{3}{2}\sum_{i=1,2}\left[C_i-\frac{C_i-8Z_i}{5}I\right]\left(\frac{\Lambda_i}{k_F^0}\right)^3
\left(\frac{\left((1-I)u\right)^{1/3}}{\frac{\Lambda_i}{k_F^0}}-
\tan^{-1}
\frac{\left((1-I)u\right)^{1/3}}{\frac{\Lambda_i}{k_F^0}}\right)
\nonumber.
\end{eqnarray}
In Eq.~(\ref{e-T0}), $I$ is the asymmetry parameter
($I=(n_n-n_p)/n$) and $u=n/n_0$, with $n_0$ denoting the
equilibrium symmetric nuclear matter density, $n_0=0.16$
fm$^{-3}$. The parameters $A$, $B$, $\sigma$, $C_1$, $C_2$ and
$B'$ which appear in the description of symmetric nuclear matter
are determined in order that $E(n=n_0)-mc^2=-16$ {\rm MeV},
$n_0=0.16$ fm$^{-3}$, and the incompressibility is $K=240$ {\rm
MeV} and have the values $A=-46.65$, $B=39.45$, $\sigma=1.663$,
$C_1=-83.84$, $C_2=23$ and $B'=0.3$. The finite range parameters
are $\Lambda_1=1.5 k_F^{0}$ and $\Lambda_2=3 k_F^{0}$ and $k_F^0$
is the Fermi momentum at the saturation point $n_0$.

 \begin{table}[h]
 \begin{center}
\caption{The parameters for neutron rich matter of MDI model. }
 \label{t:1}
\vspace{0.5cm}
\begin{tabular}{|c|cccc|}
\hline
 $F(u)$  &   $x_0$    &    $x_3$    & $Z_1$ &
$Z_3$      \\
\hline
 $\sqrt{u}$        & 0.376  & 0.246 & -12.23 & -2.98   \\
 \hline
 $u$              &  0.927  & -0.227  & -11.51& 8.38     \\
 \hline
 $2u^2/(1+u)$      & 1.654  & -1.112 & 3.81 & 13.16 \\
 \hline
\end{tabular}
\end{center}
\end{table}


The additional parameters $x_0$, $x_3$, $Z_1$, and $Z_2$ employed
to determine the properties of asymmetric nuclear matter are
treated as parameters constrained by empirical knowledge
\cite{Prakash-97} and presented in Table~1. By suitably choosing
the parameters $x_0$, $x_3$, $Z_1$, and $Z_2$, it is possible to
obtain different forms for the density dependence of the symmetry
energy $E_{sym}(u)$. The nuclear symmetry energy is parametrized
according to the following formula
\begin{equation}
E_{sym}(n)= 13 u^{2/3}+17 F(u),\label{Esym-3}
\end{equation}
where the first term of the right-hand side of Eq.~(\ref{Esym-3})
represents   the contribution of the kinetic energy and the second
term is the contribution of the interaction energy. In general, in
order to obtain different forms for the density dependence of
$E_{sym}(n)$, the function $F(u)$  can be parameterized as follows
\cite{Prakash-97}
\begin{equation}
F(u)=\sqrt{u},\qquad  F(u)=u, \qquad F(u)=2u^2/(1+u). \label{fu-1}
\end{equation}
In the present work its parametrization corresponds to the models
MDI-1, MDI-2 and MDI-3 respectively.  Numerical values of the
parameters that generate these functional forms are given in Table
1. It is worthwhile to point out that the above parametrization of
the interaction part of the nuclear symmetry energy is extensively
used for the study of neutron star properties
\cite{Prakash-97,Prakash-94} as well as the study of the
collisions of neutron-rich heavy ions at intermediate energies
\cite{Li-97,Baran-05}.

The pressure of the baryons, at $T=0$,  defined as
\begin{equation}
P=n^2\frac{\partial E(n,I)}{\partial n}, \label{Pres-def-1}
\end{equation}
with the help of Eq.~(\ref{e-T0}) takes the analytical form
\begin{eqnarray}
P(n,I)&=&\frac{1}{5}n_0E_F^0
u^{5/3}\left[(1+I)^{5/3}+(1-I)^{5/3}\right]+\frac{1}{3}n_0u^2A\left[\frac{3}{2}-\left(\frac{1}{2}+x_0\right)I^2\right]
\nonumber \\
\\
&+&\frac{2}{3}B\sigma n_0 u^{\sigma
+1}\frac{\left[\frac{3}{2}-(\frac{1}{2}+x_3)I^2\right]\left(1+\frac{2}{3\sigma}B'u^{\sigma
-1}\left[\frac{3}{2}-(\frac{1}{2}+x_3)I^2\right]
\right)}{\left(1+\frac{2}{3}B'\left[\frac{3}{2}-(\frac{1}{2}+x_3)I^2\right]u^{\sigma-1}\right)^2}
\nonumber \\
&+&
\frac{n_0u^2}{2}\sum_{i=1,2}\left[C_i+\frac{C_i-8Z_i}{5}I\right]
\left(\frac{\Lambda_i}{k_F^0}\right)^2
\frac{(1+I)^{1/3}}{u^{2/3}}\left(1-\frac{1}{1+\frac{(1+I)^{2/3}u^{2/3}}{\left(\frac{\Lambda_i}{k_F^0}\right)^2}}\right)
\nonumber \\
&+&\frac{n_0u^2}{2}\sum_{i=1,2}\left[C_i-\frac{C_i-8Z_i}{5}I\right]
\left(\frac{\Lambda_i}{k_F^0}\right)^2
\frac{(1-I)^{1/3}}{u^{2/3}}\left(1-\frac{1}{1+\frac{(1-I)^{2/3}u^{2/3}}{\left(\frac{\Lambda_i}{k_F^0}\right)^2}}\right).
\nonumber \label{P-all-T0}
\end{eqnarray}

\subsection{Skyrme  model}
The Skyrme functional providing the energy per baryon of
asymmetric nuclear matter is given by the formula
\cite{Chabanat-97,Farine-97}
\begin{eqnarray}
E(n,I)&=&\frac{3}{10}\frac{\hbar
^2c^2}{m}\left(\frac{3\pi^2}{2}\right)^{2/3}n^{2/3}F_{5/3}(I)+\frac{1}{8}t_0n\left[2(x_0+2)-(2x_0+1)F_2(I)\right]
\nonumber \\
&+&\frac{1}{48}t_3n^{\sigma+1}\left[2(x_3+2)-(2x_3+1)F_2(I)\right] \\
&+& \frac{3}{40}\left(\frac{3\pi^2}{2}\right)^{2/3}n^{5/3}
\left[\frac{}{}\left(t_1(x_1+2)+t_2(x_2+2)\right)F_{5/3}(I)
\right.
\nonumber\\
&+&\left.\frac{1}{2}\left(t_2(2x_2+1)-t_1(2x_1+1)\right)F_{8/3}(I)\right],
\nonumber
\end{eqnarray}
where $\displaystyle F_m(I)=\frac{1}{2}\left[(1+I)^m+(1-I)^m
\right] $ and the parametrization is given in
Refs~\cite{Chabanat-97,Farine-97}.

\subsection{Thomas-Fermi model}
The model's prediction for the equation of state reads $E(n,I)=T_0
\eta(u,I)$, where \cite{Myers-98}
\begin{equation}
\eta(u,I)=a(I)\Omega^3-b(I)\Omega^3+c(I)\Omega^5, \quad
\Omega\equiv(n/n_0)^{1/3}, \quad n_0=0.16114 \ {\rm fm}^{-3}
\label{TF-ap-1}
\end{equation}
$n_0$ and $T_0$ are the saturation density and Fermi energy of
standard nuclear matter accordingly as predicted by the model and
the coefficients $a(I)$, $b(I)$, $c(I)$ are the following
functions of $I$
\begin{equation}
a(I)=\frac{3}{20}\left[2(1-\gamma_l)(p^5+q^5)-\gamma_u \left\{
\begin{array}{ll}
(5p^2q^3-q^5)  & {\rm for}\quad  n_n \geq n_p            \\
(5p^3q^2-p^5)  & {\rm for}\quad   n_n \leq n_p       \\
                              \end{array}
                       \right.
\right], \label{TF-ap-2}
\end{equation}
\begin{equation}
b(I)=\frac{1}{4}\left[\alpha_l(p^6+q^6)+2\alpha_up^3q^3  \right],
\label{TF-ap-3}
\end{equation}
\begin{equation}
c(I)=\frac{3}{10}\left[B_l(p^8+q^8)+B_up^3q^3(p^2+q^2) \right],
\label{TF-ap-4}
\end{equation}
where $p=(1+I)^{1/3}, \ q=(1-I)^{1/3}$. The interaction strengths
$\gamma_l, \gamma_u, \alpha_l, \alpha_u, B_l,B_u$ have the
following values: $\gamma_l=0.25198$,  $\gamma_u=0.88474$,
$\alpha_l=0.7011$, $\alpha_u=1.24574$, $B_l=0.22791$,
$B_u=0.8002$.

\section{Results and Discussion}
\begin{figure}
\centering
\includegraphics[height=6.1cm,width=5.1cm]{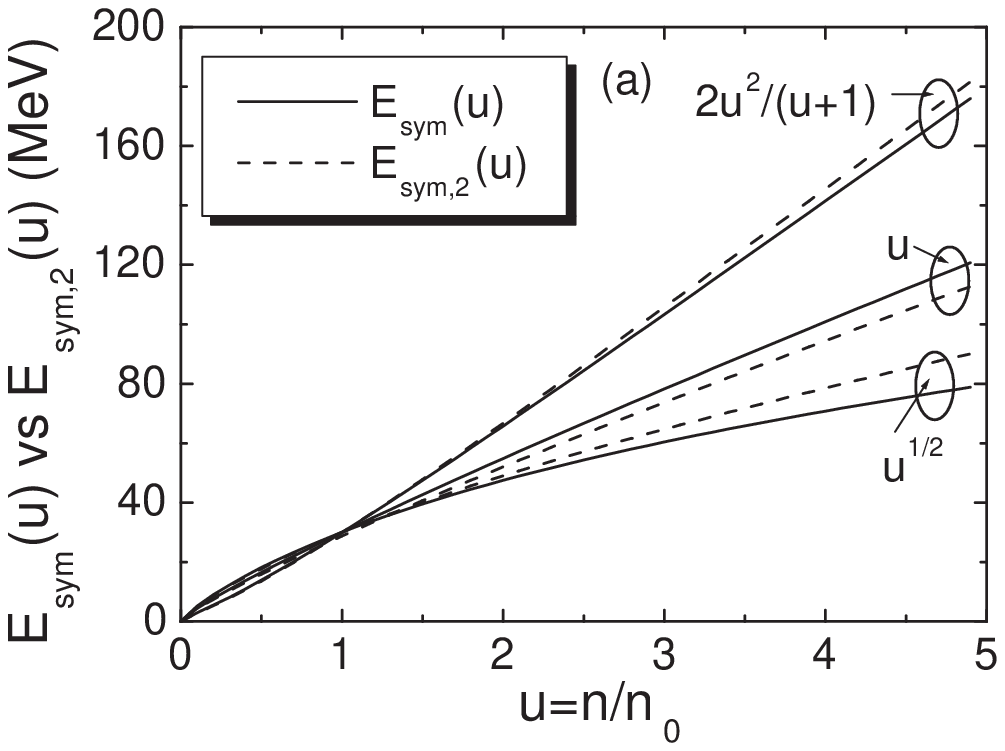}\
\includegraphics[height=6.1cm,width=5.1cm]{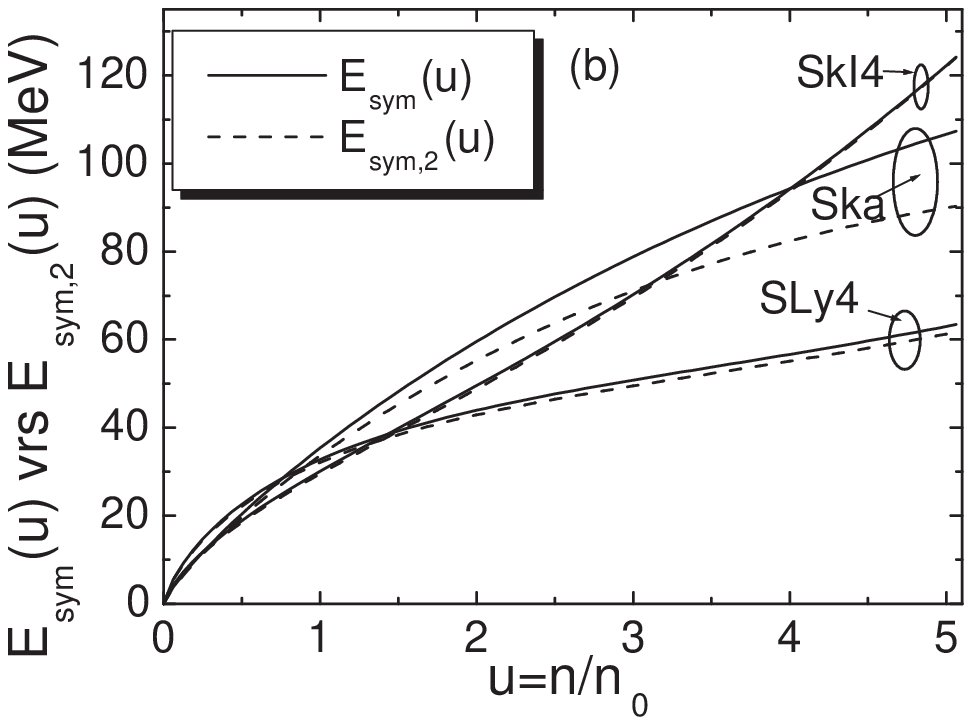}\
\includegraphics[height=6.1cm,width=5.1cm]{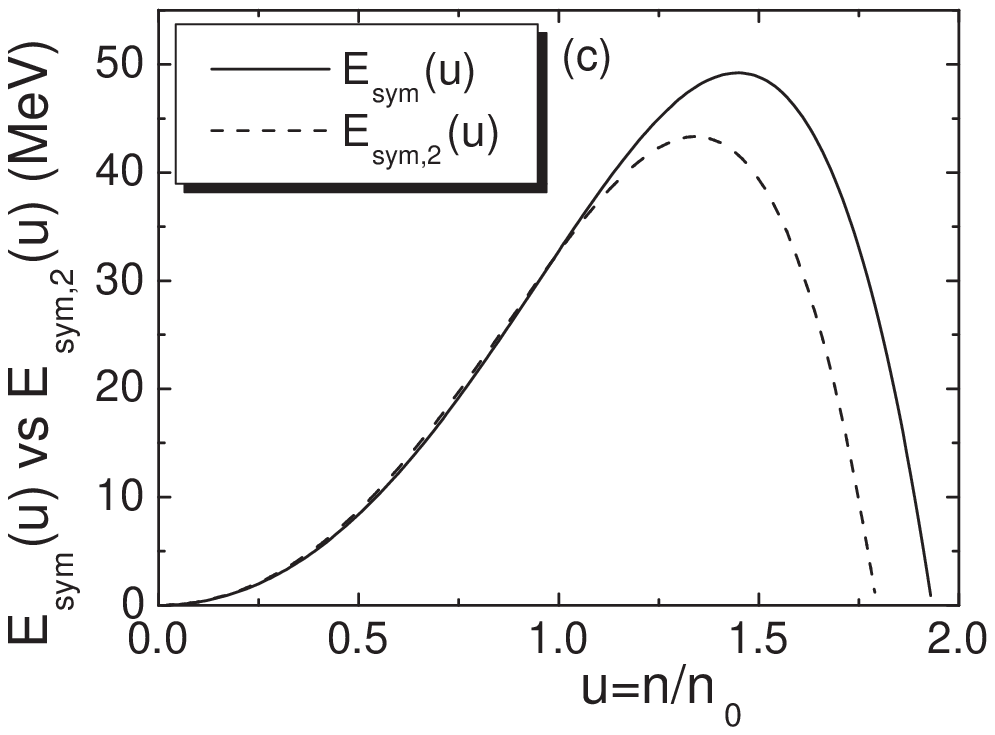}\
\caption{The symmetry energy $E_{sym}(u)$ defined by
Eq.~(\ref{esym-def}) versus the symmetry energy $E_{sym,2}(u)$
defined by Eq.~(\ref{Expan-2}) as a function of the density
fraction $u=n/n_0$ for the three considered models. } \label{}
\end{figure}
\begin{figure}
\centering
\includegraphics[height=6.1cm,width=5.1cm]{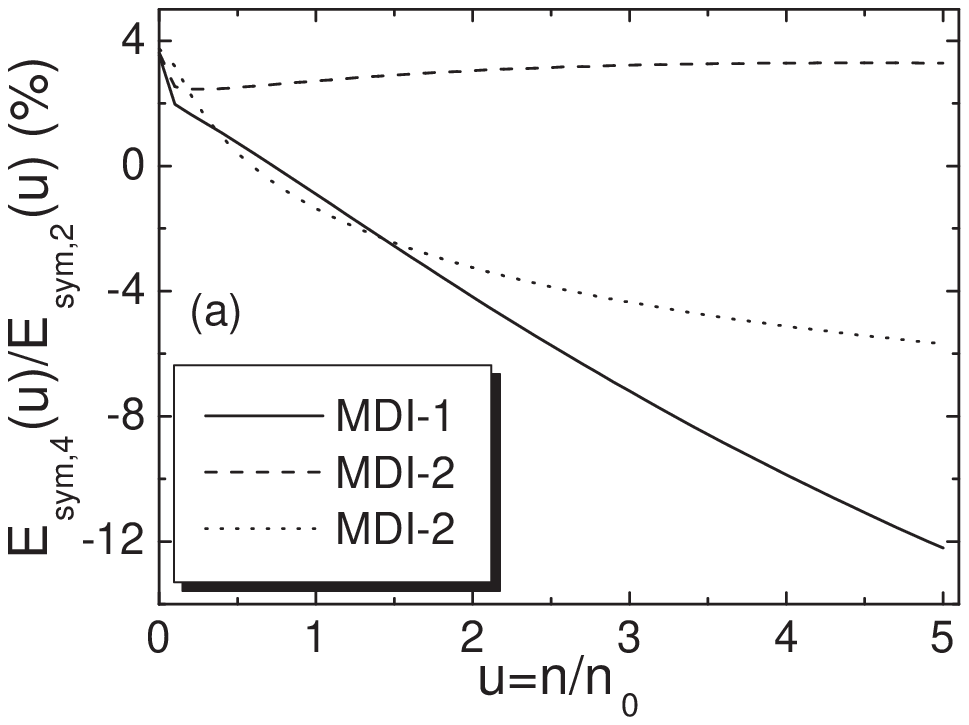}\
\includegraphics[height=6.1cm,width=5.1cm]{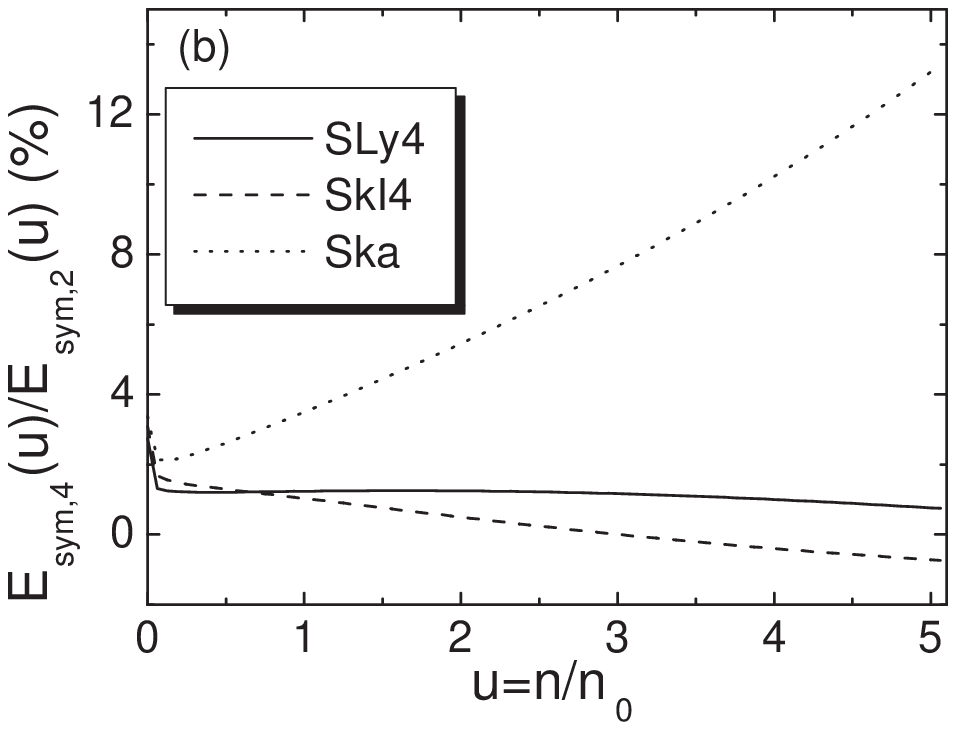}\
\includegraphics[height=6.1cm,width=5.1cm]{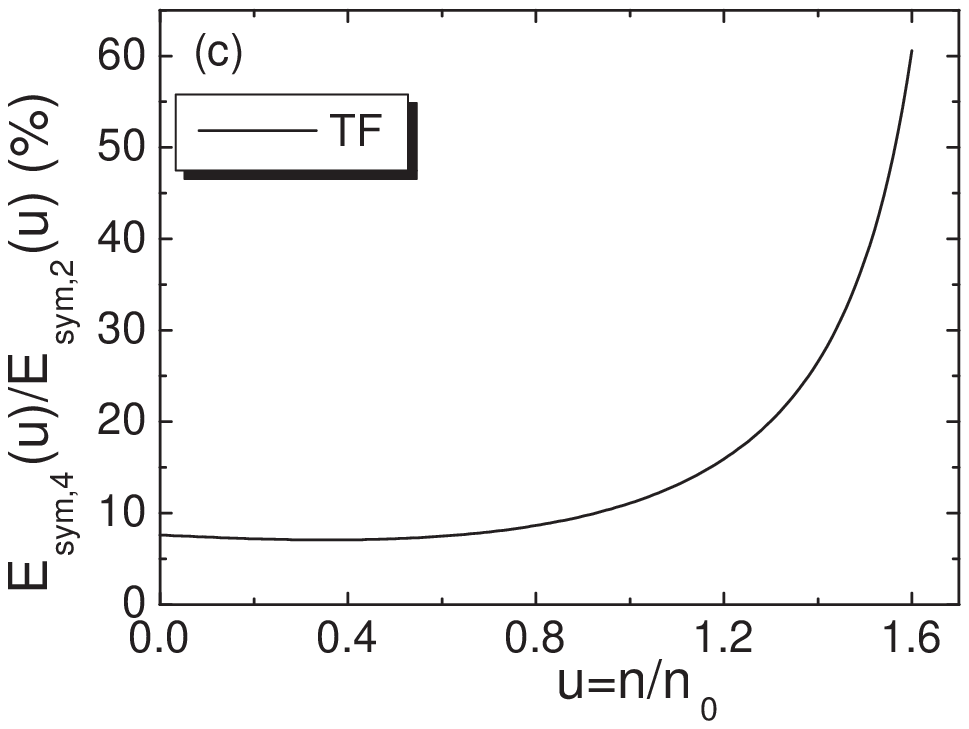}\
\caption{The density dependence of the ratio
$E_{sym,4}(u)/E_{sym,2}(u)$ for the three considered models.}
\label{}
\end{figure}
\begin{figure}
\centering
\includegraphics[height=6.1cm,width=5.1cm]{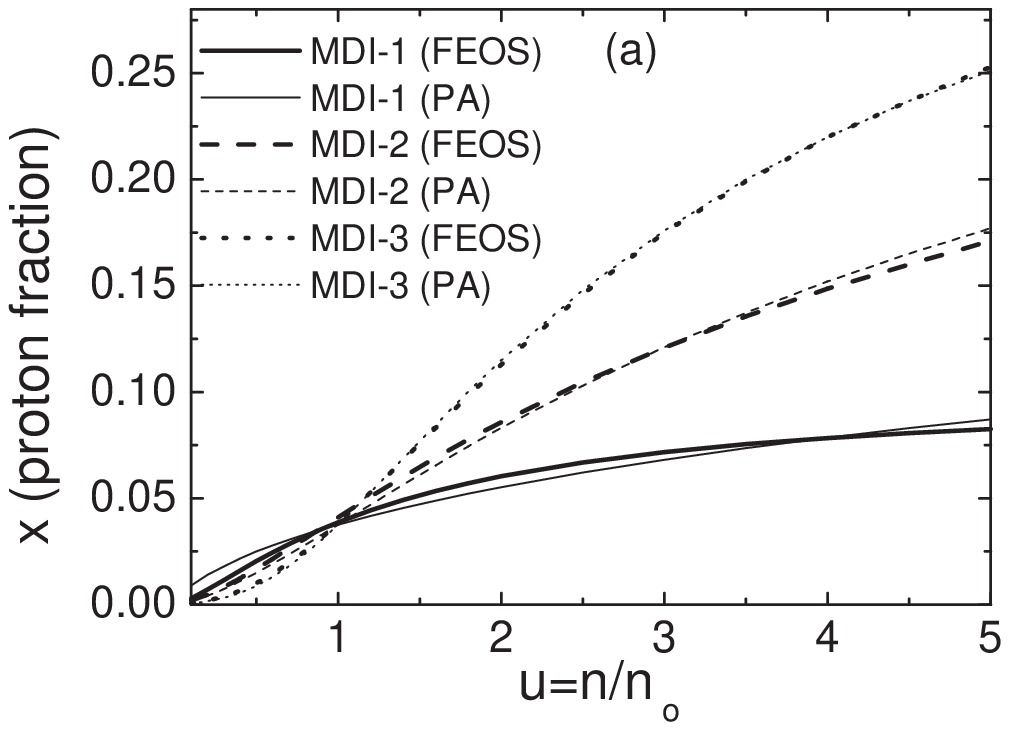}\
\includegraphics[height=6.1cm,width=5.1cm]{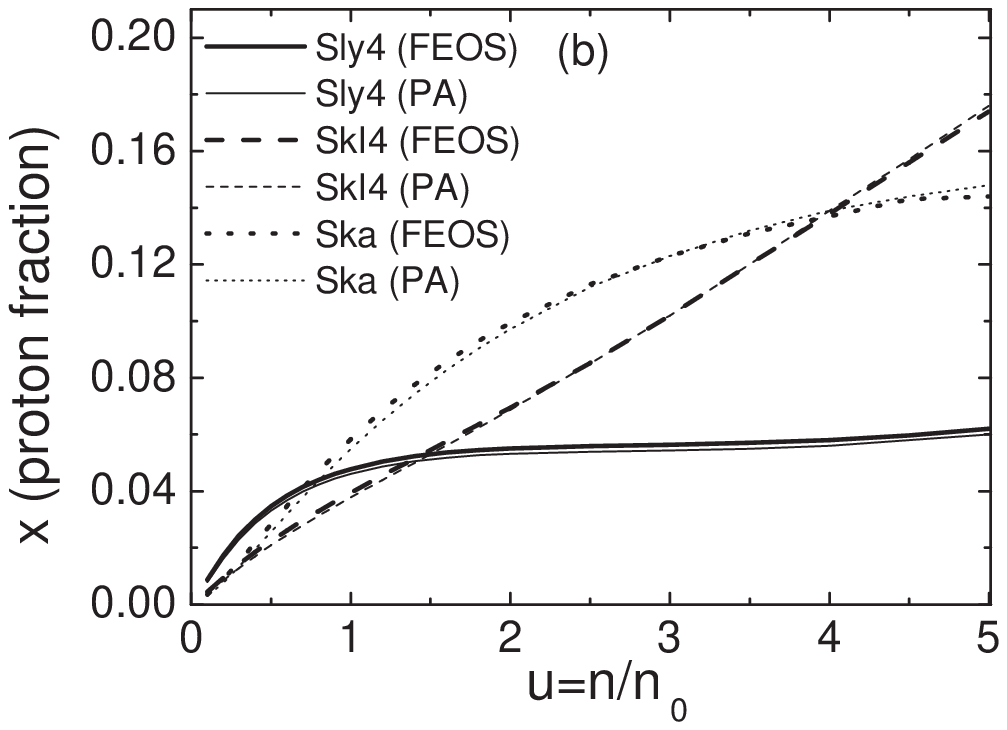}\
\includegraphics[height=6.1cm,width=5.1cm]{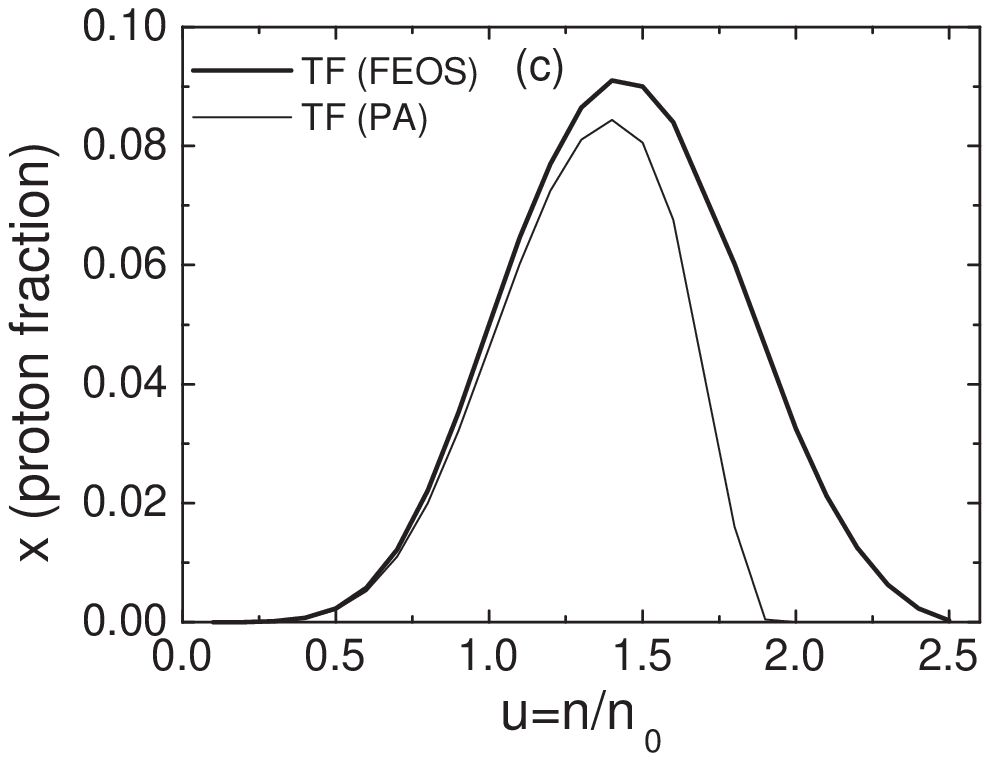}\
\caption{The density dependence of the proton fraction $x$
obtained from the considered models   employing the FEOS as well
as the PA. } \label{}
\end{figure}

In Fig.~1 we compare the density dependence of the quantities
$\displaystyle E_{sym,2}(n)=\left.
\frac{1}{2!}\frac{\partial^2E(n,I)}{\partial I^2} \right|_{I=0}$
and $E_{sym}(n)=E(n,I=1)-E(n,I=0)$. Both definitions of the
symmetry energy exhibit a similar behavior for low values of the
baryon density (up to $u\simeq 1.5$). The difference, which is
model dependent, is more pronounced for higher values of $n$. This
is a direct indication that the two approaches affect the equation
of state of high density nuclear matter at the inner core of
neutron stars. It is worth noticing that for the three MDI models
and the three Skyrme models the symmetry energy is an increasing
function of the density. In TF model the symmetry energy is
increasing an function of $u$ only up to $u\simeq 1.5$ and then
decreases rapidly  with $u$. This has a dramatic effect on neutron
stars applications mainly on those connected with the high density
equation of state.

In order to check the accuracy of the parabolic approximation we
display  in  Fig.~2 the density dependence of the ratio
$E_{sym,4}(u)/E_{sym,2}(u)$. It is seen that in  most cases the
contribution of the fourth-order term $E_{sym,4}(u)$ is less that
$4 \%$ compared to the second order one $E_{sym,2}(u)$, at least
for low values of the density ($u \leq 2$). However, in four of
the considered models, i.e. the MDI-1, MDI-3, the SKa and TF the
contribution of $E_{sym,4}(u)$ increases rapidly with the density
and consequently  influences the high density equation of state
and this must be taken into account. In the specific case of the
TF model the contribution of $E_{sym,4}(u)$ becomes comparable to
$E_{sym,2}(u)$ even for low values of $u$, making the parabolic
approximation problematic.

In Fig.~3 we compare the density dependence of proton fraction $x$
in $\beta$-stable nuclear matter determined by employing the full
EOS (FEOS)  and the parabolic approximation. In the case of the
MDI and Skyrme models there is a difference  (depended on the
specific model) mainly between 4 \% and 20\% for low values of the
baryon density (up to $u \simeq 0.7$). However, the parabolic
approximation is good for higher values of $n$. We expect that the
above density dependence of $x$ will be reflected also on the
values of the transition density $n_t$ and pressure $P_t$. In the
case of the TF model only for low values of the density ($u \leq
1$) the two approximation produce similar results. Moreover, it is
concluded that the parabolic approximation does not affect
appreciably the onset of the URCA process with critical values
$x_{Urca}\simeq 11 \% $ (without muons) and $x_{Urca}\simeq 14 \%$
by including muons.
\begin{figure}
\centering
\includegraphics[height=6.2cm,width=6.2cm]{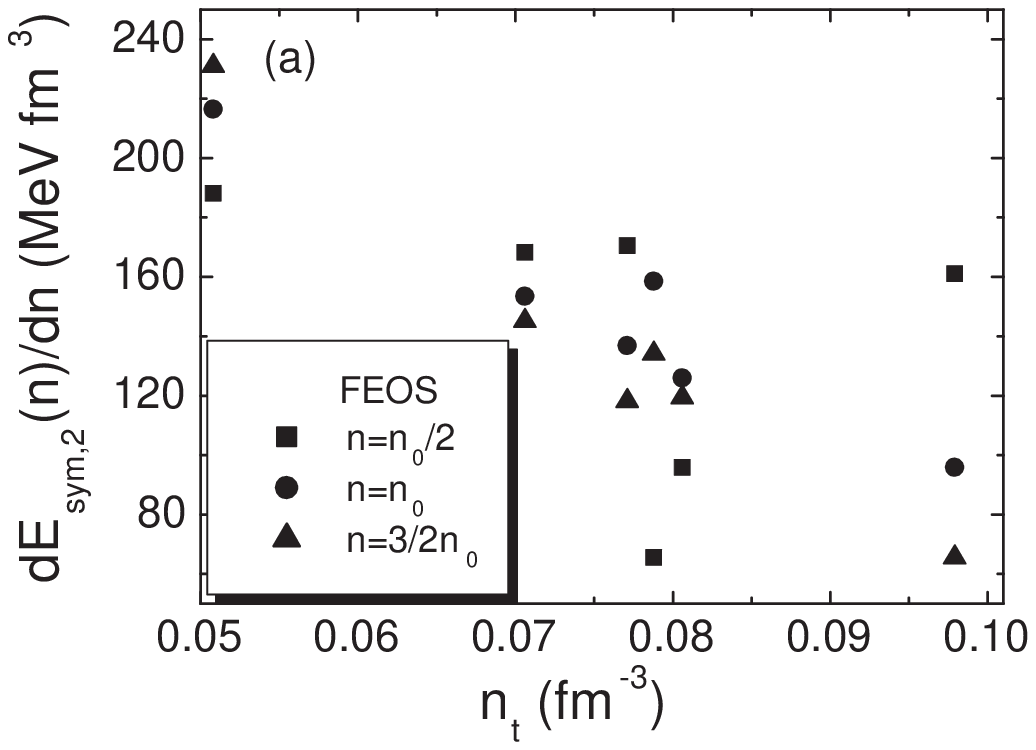}\
\includegraphics[height=6.2cm,width=6.2cm]{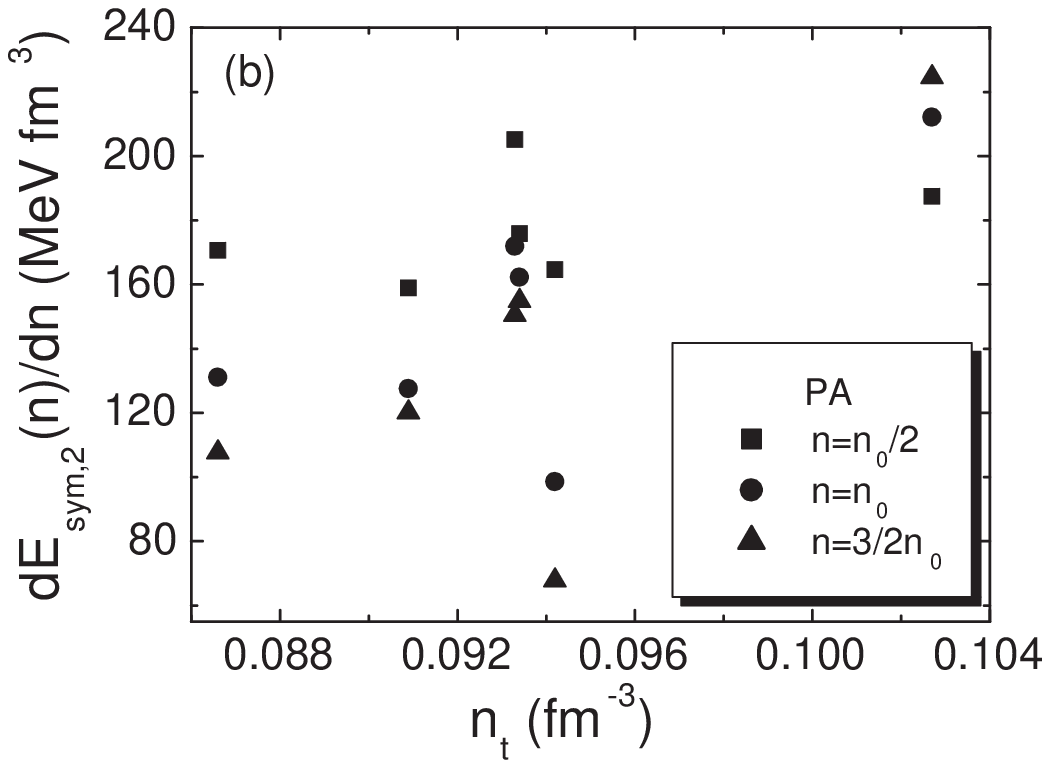}\
\caption{The derivative of the symmetry energy $dE_{sym,2}(n)/dn$
versus the transition density $n_t$ corresponding  to the
considered models a) employing the full EOS and b) employing the
parabolic approximation.} \label{}
\end{figure}

\begin{figure}
\centering
\includegraphics[height=6.2cm,width=6.2cm]{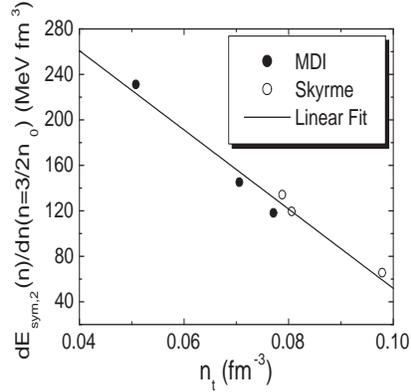}\
\caption{ The derivative of the symmetry energy
$dE_{sym,2}(n=3/2n_0)/dn$ versus the transition density, for the
FEOS, corresponding  to the considered models. The solid line
corresponds to the least-squares fit (FIT) expression
(\ref{Esym-n-1}). } \label{}
\end{figure}

In Fig.~4(a), we present the correlation between the derivative of
the symmetry energy $E_{sym,2}(n)$ (for $n=1/2n_0$, $n=n_0$ and
$n=3/2n_0$) for the MDI and Skyrme models   and the transition
density $n_t$ in the case of the FEOS. The most distinctive
feature is the concentration of the data from the case $n=1/2n_0$
(where they are rather random) to an almost linear relation in the
case $n=n_0$  and linear  in the case $n=3/2n_0$. To further
illustrate this point, we plot in Fig.~5 the derivative of the
symmetry energy $E'_{sym,2}(n)$ at the baryon density $n=3n_0/2$
versus $n_t$. A linear relation is found of the form
\begin{equation}
E'_{sym,2}\left(n=\frac{3n_0}{2}\right)=400.18-3483.3 n_t, \qquad
({\rm MeV} \cdot {\rm fm}^3). \label{Esym-n-1}
\end{equation}
%
%
We find that the values of $n_t$ depend on the trend of the
equation of state and the symmetry energy depends not only on
values close to the saturation density $n_0$ but even on higher
ones. It is of interest to see that, according to Fig.~4(b) in the
case of the PA the data are scattered even for values close to
$n=3/2n_0$. It is presumed that the use of the PA may affect the
linear correlation between $E'_{sym,2}(n)$ and $n_t$ for
$n=3/2n_0$.

 \begin{table}[h]
 \begin{center}
\caption{The transition density $n_t$ (in fm$^{-3}$) and pressure
$P_t$ (in MeV) obtained from the considered models  by employing
the FEOS as well as the parabolic approximation. } \label{t:2}
\vspace{0.5cm}
\begin{tabular}{|c|ccccccc|}
\hline
 approach &MDI-1 & MDI-2 & MDI-3    & TF & Sly4 &  SKI4 & Ska       \\
\hline
$n_{t}$ (FEOS)       & 0.0771      & 0.0706    & 0.0508    &  0.1368   & 0.0979     & 0.0806      & 0.0788    \\
\hline
$n_{t}$ (PA)  & 0.0866      & 0.0934    & 0.1027    &  0.1419   &  0.0942    &  0.0909    &  0.0933   \\
\hline
$P_t$(FEOS)          & 0.3456      & 0.2338    & 0.3133    &   2.9020  &  0.5781    &  0.3365    & 0.5295      \\
\hline
$P_t$(PA)     & 0.2975      & 0.7055    & 1.2482    &   3.7300  &  0.5459   &  0.4965    &   0.8651    \\
\hline
\end{tabular}
\end{center}
\end{table}

In Table~2 we present the values of $n_t$ and $P_t$ determined by
employing the FEOS and the parabolic approximation for the models
considered  in the present work. In most of the cases (the only
exception is the  Sly4 case) the use of the PA increases the
values of $n_t$ by 10-15 \% or even more (see the case MDI-3). The
effect of the PA is even more dramatic in the case of $P_t$. In
most of the cases $P_t$ increases significantly compared to the
FEOS (the only exceptions are the MDI-1 and the Sly4 models). The
increase is even two or three times. In order to clarify further
this point we plot in Fig.~6 the values of  $P_t$ versus $n_t$ for
the considered models. We see the strong dependence of $P_t$ on
$n_t$ in the case of PA compared to FEOS. The above results
indicate that one may introduce a large error by employing the
parabolic approximation in order to determine the value of $P_t$.

\begin{figure}
\centering
\includegraphics[height=6.2cm,width=6.2cm]{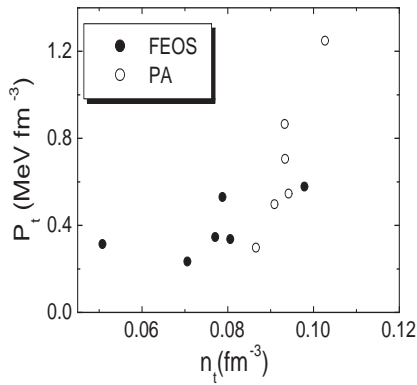}\
\caption{ The transition pressure $P_t$ versus the transition
density $n_t$ employing the full and the parabolic approximation.
} \label{}
\end{figure}

The values of $P_t$, according to Eq.~(\ref{inertia-1}), strongly
influence the value of the crustal fraction of the moment of
inertia. Actually, the crustal fraction $\Delta I/I$ depends not
only on $P_t$ and $n_t$ but on the mass $M$ and radius of $R$ of
the neutron star as well. Actually $M$ and $R$ depend also on
$P_t$ and $n_t$. In order to have a qualitative picture of the
effect of $P_t$ and $n_t$ on $\Delta I/I$ we can consider the
maximum mass and the corresponding radius $R$ of the neutron star
coming from the MDI model~\cite{Prakash-88} and neglecting  the
effects of $P_t$ and $n_t$ (which is a good approximation).  We
determine the ratio $\Delta I/I$ by considering the full EOS and
the parabolic approximation. The results are presented in Table~3.
Its obvious that there is a strong difference  between of two
approaches. For example in the case of MDI-2 the PA increases more
than two times the value of $\Delta I/I$ compared to the FEOS,
while in the case of MDI-3 the increase is even higher.

 \begin{table}[h]
 \begin{center}
\caption{The crustal fraction of the moment of inertia for the
three MDI models   employing the FEOS as well as the parabolic
approximation.} \label{t:3} \vspace{0.5cm}
\begin{tabular}{|c|cccc|}
\hline
 Model  &   $M_{max}/M_{\odot}$    &    $R (Km)$    & $\Delta I/I$(FEOS) &
$\Delta I/I$(PA)      \\
\hline
 MDI-1        & 1.895  & 10.112 & 0.0076 & 0.0068   \\
 \hline
 MDI-2              &  1.935  & 10.570  & 0.0062& 0.0163     \\
 \hline
 MDI-3      & 1.952  & 10.933 & 0.0086 & 0.0287 \\
 \hline
\end{tabular}
\end{center}
\end{table}


The limit $\Delta I/I=0.014$ constrained by Link {\it et
al}.~\cite{Link-99} limits also the masses and the radii of the
neutron star for  specific values of $P_t$ and $n_t$. It is
concluded that the strong dependence  of $\Delta I/I=0.014$ on
$P_t$ puts also strong constraints  on the allowed pairs of $M$
and $R$. In Fig.~7 we plot the M-R constrained relation for the
Vela pulsar where $\Delta I/I >0.014$ obtained from the full and
parabolic approximation in the case of  the MDI model and  the
Skyrme model. It is obvious that the implication of the full EOS
imposes more restrictive constraints compared to the parabolic
one. This effect is more pronounced in the case of the MDI models
compared to Skyrme model. Consequently, the use of the PA may also
introduce a large error in the determination of the minimum radius
(for a fixed value of a mass) of a neutron star.

In Fig.~8 we compare the r-mode instability window, obtained from
the FEOS and PA in the case of  the MDI model, with those of the
observed neutron stars in low-mass x-ray binaries (LMXBs) for
$M=1.4 M_{\odot}$. The critical frequencies are strongly localized
at high values  in the case of the PA. The employment of the FEOS
drops significantly the instability window, especially in the case
of the stiffer equation of state (MDI-3 case). More specifically,
the employment of the FEOS comparing to PA decreases the values of
the  critical frequency $\nu_c$  around 4$\%$ (MDI-1), 8$\%$
(MDI-2) and 24$\%$ (MDI-3).

In addition,  following the study of Wen {\it et
al.}~\cite{Wen-012} we examine four cases of LMXBs that is the 4U
1608-522 at 620 Hz, 4U 1636-536 at 581 Hz, MXB 1658-298 at 567 Hz
and EXO 0748-676 at 552 Hz \cite{Watts-08,Keek-010}. The masses of
the mentioned stars are not measured accurately but the core
temperature $T$ is derived from their observed accretion
luminosity. It is obvious from Fig.~8 that for a $M=1.4 M_{\odot}$
three  of the considered LMXBs lie inside  instability window.
According to discussion of Ref.~\cite{Wen-012} and finding in
Refs.~\cite{Levin-99,Bondarescu-07} the LMXBs should be out of the
instability window. Consequently, one can presume that either the
LMXBs masses are even lower than  $M=1.4 M_{\odot}$ or the softer
equation of state is more preferred. However, additional
theoretical and observation work  must be dedicated before a
definite  conclusion.
\begin{figure}
 \centering
\includegraphics[height=7.5cm,width=7.5cm]{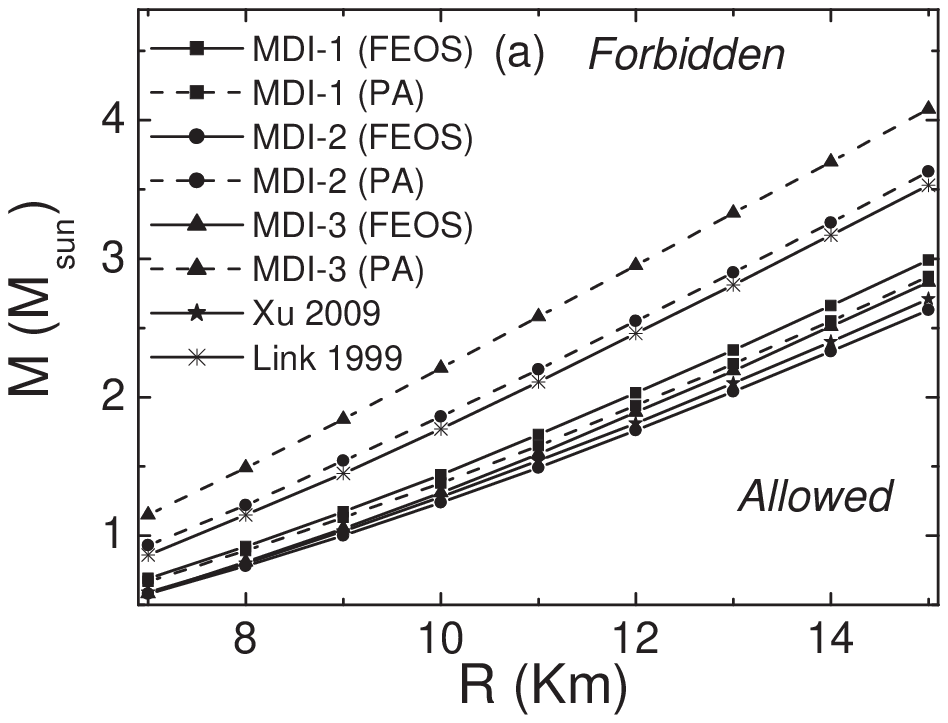}\
\includegraphics[height=7.5cm,width=7.5cm]{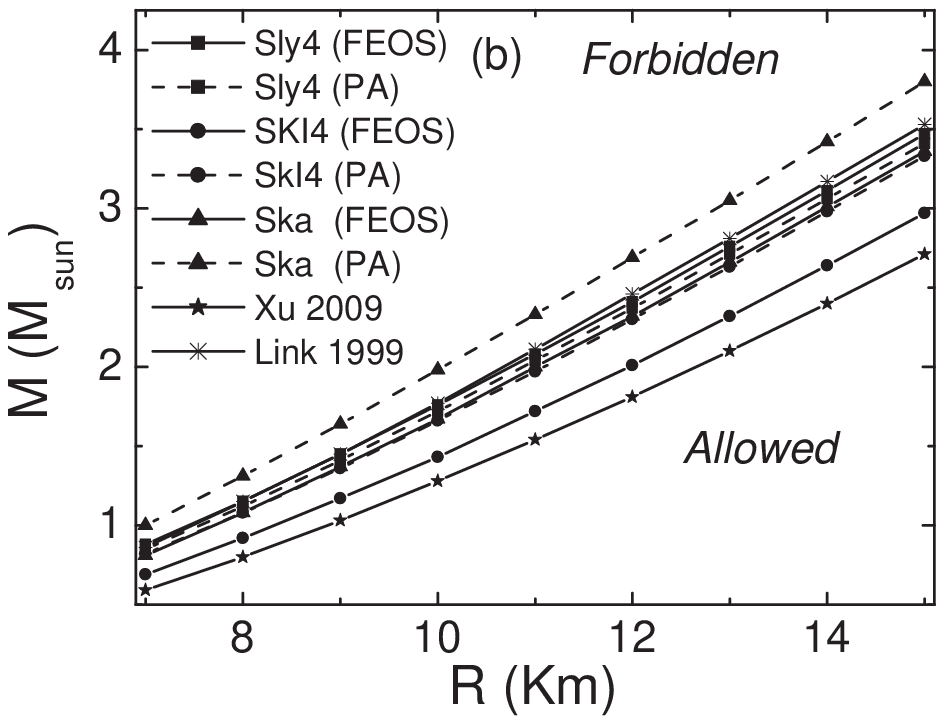}\
\caption{The M-R constrained relation for the Vela pulsar where
$\Delta I/I >0.014$ obtained from the FEOS and PA in the case of
a) the MDI model and b) the Skyrme model. The constraint implies
that allowed masses (in $M_{\odot}$) and radii lie to the right of
the line. The two additional constraints are taken from Xu {\it et
al.}~\cite{Xu-09-1} ($n_t=0.065$ fm$^{-3}$ and $P_t=0.26$MeV
fm$^{-3}$) and Link {\it et al.}~\cite{Link-99} ($n_t=0.075$
fm$^{-3}$ and $P_t=0.65$MeV fm$^{-3}$). } \label{}
\end{figure}

\begin{figure}
\vspace{2cm} \centering
\includegraphics[height=7cm,width=8cm]{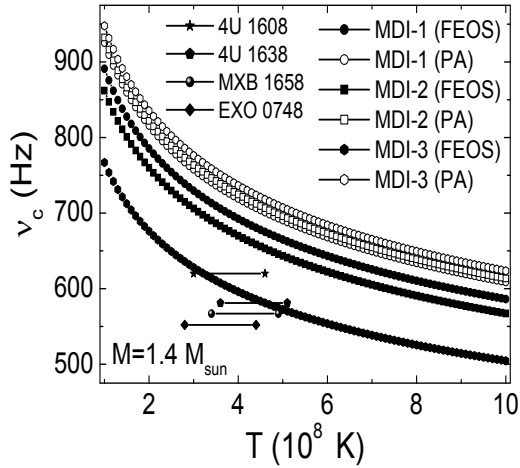}
\vspace{-3cm} \caption{ The critical frequency  temperature
dependence for a neutron star with mass $M=1.4 M_{\odot}$ obtained
from the FEOS and PA in the case of  the MDI model. In addition,
the location of the observed short-recurrence-time LMXBs
\cite{Watts-08,Keek-010}.}  \label{}
\end{figure}

\section{Summary}
The study of symmetry energy effects on location of the inner edge
of neutron star crusts is the main topic of the present work. We
employ three different phenomenological models for the prediction
of the EOS and we determine in the framework of thermodynamical
method both the transition density and transition pressure
corresponding to the inner edge of a neutron star crust. Actually,
$n_t$ and $P_t$ have been determined separately by employing the
full equation of state of asymmetric matter and its parabolic
approximation. Our results confirm the statement that  employing
the PA instead of the full EOS may influence  appreciably the
values of $n_t$ and even more the values of $P_t$. We found that
there is a linear correlation between the derivative of the
symmetry energy $E'_{sym,2}(n)$ at the baryon density $n=3n_0/2$
and the transition density $n_t$ for the considered models. We
found also that the implication of the full EOS imposes more
restrictive constraints, on  the M-R  constrained relation for the
Vela pulsar, compared to the parabolic one.  Furthermore, the
employment of the FEOS drops significantly the instability window,
especially in the case of the stiff equation of state.
Consequently, the error introduced by assuming that EOS is
parabolic may induce also a large error in the determination of
related properties of a neutron star as the crustal fraction on
the moment of inertia  and the critical frequency of rotating
neutron stars.

\section*{Acknowledgments}
This work was supported by the German Science Council (DFG) via
SFB/TR7. The author would like to thank the Theoretical
Astrophysics Department of the University of Tuebingen, where part
of this work was performed  and Professor K. Kokkotas  for his
useful comments on the preparation of the manuscript. The author
thanks Dr. C.P. Panos for his remarks on the present paper.




\begin{thebibliography}{99}
%
\bibitem{Shapiro-83}S.L. Shapiro and S.A. Teukolsky, {\it Black
Holes, White Dwarfs, and Neutron Stars} (John Wiley and Sons, New
York, 1983).
%
\bibitem{Haensel-07}P. Haensel, A.Y. Potekhin, and D.G. Yakovlev,
{\it Neutron Stars 1: Equation of State and Structure}
(Springer-Verlag, New York, 2007).
%
\bibitem{Lattimer-07} J.M. Lattimer and M. Prakash, Phys. Rep.
{\bf 442}, 109 (2007).
%
%
\bibitem{Pethick-95b} C.J. Pethick, D.G. Ravenhall, and C.P.
Lorenz,  Nucl. Phys. A {\bf 584}, 675 (1995).
%
\bibitem{Baym-71a}G. Baym, C. Pethick, and P. Sutherland, Astrophys. J. {\bf 170}, 299 (1971).
%
\bibitem{Baym-71b}G. Baym, H. A. Bethe, and C.J. Pethick, Nucl. Phys. A {\bf 175}, 225 (1971).
%
\bibitem{Pethick-95a}C.J. Pethick and D.G. Ravenhall, Ann. Rev.
Nucl. Part. Sci. {\bf 45}, 429 (1995).
%
\bibitem{Douchin-00}F. Douchin and P. Haensel, Phys. Let. B {\bf
485},  107 (200).
%
\bibitem{Oyamatsu-07}K. Oyamatsu and K. Iida,  Phys. Rev. C {\bf 75}, 015801 (2007).
%
\bibitem{Ducoin-07}C. Ducoin, Ph. Chomaz, and F. Gulminelli,  Nucl. Phys. A {\bf 789}, 403 (2007).
%
\bibitem{Xu-09-1}J. Xu, L.W. Chen, B.A. Li, and H.R. Ma, Phys. Rev. C {\bf 79}, 035802 (2009).
%
\bibitem{Xu-09-2}J. Xu, L.W. Chen, B.A. Li, and H.R. Ma, Astrophys. J. {\bf 697}, 1549 (2009).
\bibitem{Kubis-07}S. Kubis, Phys. Rev. C {\bf 76}, 025801 (2007).
%
\bibitem{Worley-08}A. Worley, P.G. Krastev, and B.A. Li, Astrophys. J. {\bf 685}, 390 (2008).
%
\bibitem{Kubis-04}S. Kubis, Phys. Rev. C {\bf 70}, 065804 (2004).
%
\bibitem{Horowitz-01}C.J. Horowitz and J. Piekarewicz, Phys. Rev.
Let. {\bf 86}, 5647 (2001).
%
\bibitem{Carriere-03}J. Carriere, C.J. Horowitz, and J.
Piekarewicz, Astrophys. J. {\bf 593}, 463 (2003).
%
\bibitem{Ducoin-011}C. Ducoin, J. Margueron, C. Providencia and I.
Vidana, Phys. Rev. C {\bf 83}, 045810 (2011).
%
\bibitem{Ducoin-010}C. Ducoin, J. Margueron and C. Providencia,
Eur. Phys. Lett. {\bf 91}, 32001 (2010).
%
\bibitem{Moustakidis-010}Ch.C. Moustakidis, T. Nik\v{s}i\'{c}, G.A. Lalazissis,  D. Vretenar and P.
Ring,  Phys. Rev. C {\bf 81}, 065803 (2010).
%
%
\bibitem{Xu-010}J. Xu, L.W. Chen,  C.M. Ko, and B.A. Li,  Phys. Rev. C {\bf 81}, 055805 (2010).
%
\bibitem{Cai-012}B.J. Cai and L.W. Chen, Phys. Rev. C {\bf 85}, 024302 (2012).
%
%
\bibitem{LCK08} B.A. Li, L.W. Chen, and C.M. Ko, Phys. Rep. \textbf{464}, 113 (2008).
%
\bibitem{Tsa04} M.B. Tsang et al., Phys. Rev. Lett. \textbf{92}, 062701(2004).
%
\bibitem{Tsang-09}M.B. Tsang, Y. Zhang, P. Danielewicz, M.
Famiano, Z. Li, W.G. Lynch, and A.W. Steiner, Phys. Rev. Let. {\bf
102}, 122701 (2009).
%
%
%
\bibitem{Callen-85} H.B. Callen, Thermodynamics, Wiley, New York,
1985.
%
\bibitem{Prakash-94} M. Prakash {\it The Equation of State and Neutron
Star} lectures delivered at the Winter School held in Puri India
1994.
%
%
\bibitem{Link-99}B. Link, R.I. Epstein, and J.M. Lattimer, Phys.
Rev. Lett., {\bf 83}, 3362, (1999).
%
\bibitem{Lidblom-2000}L. Lindblom, B.J. Owen, and G. Ushomirsky, Phys. Rev. D {\bf 62}, 084030 (2000).
\bibitem{Andersson-1998}N. Andersson, Astrophy. J. {\bf 502}, 708 (1998).
%
\bibitem{Friedman-98}J.L. Friedman and S.M. Morsink, Astrophy. J. {\bf 502}, 714 (1998).
%
\bibitem{Friedman-99}J.L. Friedman and K.H. Lockitch, Prog. Theor.
Phys. Suppl. {\bf 136}, 121 (1999).
%
\bibitem{Andersson-2001}N. Andersson and K.D. Kokkotas, Int. J.
Mod. Phys. D {\bf 10}, 381 (2001).
%
\bibitem{Andersson-2003}N. Andersson, Class. Quantum Grav. {\bf
20}, R105 (2003).
%
\bibitem{Kokkotas-99}K.D. Kokkotas and N. Stergioulas, Astron.
Astrophys. {\bf 341}, 110 (1999).
%
\bibitem{Andersson-99}N. Andersson, K. Kokkotas, and B.F. Schutz,
Astrophy. J. {\bf 510}, 846 (1999).
%
%
\bibitem{Bildsten-2000}L. Bildsten and G. Ushomirsky, Astroph. J.
Lett. {\bf 529}, L33 (2000).
%
\bibitem{Wen-012}D.H. Wen, W.G. Newton, and B.A. Li, Phys. Rev. C {\bf 85}, 025801 (2012).
%
\bibitem{Vidana-012}I. Vidana, Phys. Rev. C {\bf 85}, 045808 (2012).
%
\bibitem{Alford-2012}M.G. Alford, S. Mahmoodifar, and K.
Schwenzer, Phys. Rev. D {\bf 85}, 024007 (2000).
%
\bibitem{Prakash-97}Madappa Prakash, I. Bombaci, Manju Prakash, P.J.
Ellis, J.M. Lattimer, R. Knorren, Phys. Rep. {\bf 280}, 1 (1997).
%
\bibitem{Moustakidis-07-1}V.P. Psonis, Ch.C. Moustakidis, and S.E.
Massen, Mod. Phys. Let. A {\bf 22}, 1233 (2007).
%
\bibitem{Moustakidis-07-2}Ch.C. Moustakidis, Phys. Rev. C {\bf 76}, 025805 (2007).
%
\bibitem{Moustakidis-08}Ch.C. Moustakidis, Phys. Rev. C {\bf 78}, 054323 (2008).
%
\bibitem{Moustakidis-09-1}Ch.C. Moustakidis and C.P. Panos, Phys. Rev. C {\bf 79}, 045806 (2009).
%
\bibitem{Moustakidis-09-2}Ch.C. Moustakidis, Int. J. Mod. Phys. D, {\bf 18}, 1205 (2009).
%
\bibitem{Chabanat-97} E. Chabanat, P. Bonche, P. Haensel, J. Meyer
and R. Schaeffer, Nucl. Phys. A {\bf 627}, 710 (1997).
%
\bibitem{Farine-97}M. Farine, J.M. Pearson and F. Tondeur, Nucl. Phys. A {\bf 615}, 135 (1997).
%
\bibitem{Myers-98}W.D. Myers and W.J. Swiatecki, Phys. Rev. C {\bf 57}, 3020 (1998).
%
\bibitem{Strobel-97}K. Strobel, F. Weber, M.K. Weigel and Ch.
Schaab, Int. J. Mod. Phys. E {\bf 6}, 669 (1997).
%
\bibitem{Li-97}B.A. Li, C.M. Ko and Z. Ren, Phys. Rev. Lett. {\bf
78}, 1644 (1997).
%
\bibitem{Baran-05} V. Baran, M. Colonna, V. Greco and M.Di Toro, Phys. Rep.
{\bf 410}, 335 (2005).
%
%
\bibitem{Prakash-88}M. Prakash, T.L. Ainworth and J.M. Lattimer, Phys.
Rev. Lett., {\bf 61}, 2518 (1988).
%
%
\bibitem{Watts-08}A.L. Watts, B. Krishnam, L. Bildsten, and B.F.
Schutz, Mon. Not. R. Astron. Soc. {\bf 389}, 839 (2008).
%
\bibitem{Keek-010}L. Keek, D.K. Galloway, J.J. M. in't Zand, and
A. Heger, Astrophys. J. {\bf 718}, 292 (2010).
%
\bibitem{Levin-99}Y. Levin, Astrophys. J. {\bf 517}, 328 (1999).
%
\bibitem{Bondarescu-07}R. Bondarescu, S.A. Teukolsky, and I.
Wasserman, Phys. Rev. D {\bf 76}, 064019 (2007).
%
\end{thebibliography}
\end{document}